# A New Generation of Brain-Computer Interface Based on Riemannian Geometry

**Marco Congedo*, Alexandre Barachant and Anton Andreev**

GIPSA-lab, CNRS and Grenoble University, FRANCE

11 rue des Mathématiques, Domaine universitaire - BP 46 - 38402, Grenoble, France.

*Corresponding Author

Marco.Congedo@gmail.com

tel: +33 (0)4 76 82 62 52

fax: +33 (0)4 76 57 47 90

## Abstract

Based on the cumulated experience over the past 25 years in the field of Brain-Computer Interface (BCI) we can now envision a new generation of BCI. Such BCIs will not require training; instead they will be smartly initialized using remote massive databases and will adapt to the user fast and effectively in the first minute of use. They will be reliable, robust and will maintain good performances within and across sessions. A general classification framework based on recent advances in Riemannian geometry and possessing these characteristics is presented. It applies equally well to BCI based on event-related potentials (ERP), sensorimotor (mu) rhythms and steady-state evoked potential (SSEP). The framework is very simple, both algorithmically and computationally. Due to its simplicity, its ability to learn rapidly (with little training data) and its good across-subject and across-session generalization, this strategy a very good candidate for building a new generation of BCIs, thus we hereby propose it as a benchmark method for the field.

## Introduction

Over the past 25 years the field of brain-computer interface (BCI) has grown considerably, becoming the most prominent applied research area for electroencephalography (EEG). Thanks to substantial granting by the EC in Europe and by the NIH and NSF in the USA, among others, recently there has been a striking acceleration of BCI research and applications, both for healthy users and for clinical populations (Allison et *al*., 2012; Kübler et *al*., 2001; Tan and Nijholt, 2010; Wolpaw and Wolpaw, 2012). Yet, still today it has to be admitted that "efforts to commercialize research findings have been tiepid, hampered by a general lack of robustness when translating technologies to uncontrolled environments beyond the research laboratory" (Obeid and Picone, 2013). It has also been realized that the way to reach this objective is not the enhancement of the system complexity, instead "we need to make a balance between technological advancement and practical use in a real-world situation" (Mak et *al*., 2011). While trying to combine advances from different projects it has become evident that efforts toward the standardization of EEG data format, BCI interfaces and processing tools is of paramount importance (see for example the roadmap of the coordination "Future BNCI" project, 2012[1]). For what it concerns the processing and classification algorithm, since the inception about 20 years ago of EEG inverse solutions and of diagonalization methods such as the common spatial pattern, canonical correlation analysis, independent component analysis, with the many variants of each and possible combinations, we may say there has been no further major innovation. As a matter of fact, new methods based on these tools effectively bring upon only moderate improvement and do not increase reliability in a significant way. One source of dispersion in the field is that each one of the three main BCI modalities, namely, motor-imagery (MI), steady-state evoked potentials (SSEP) and P300, is currently treated with dedicated pre-processing, signal processing and classification tools. The classification paradigm itself is also fragmented; traditionally and still today we can divide existing paradigms in two categories: those that follow a "hard machine learning" approach, and those that use "spatial filtering" to increase the signal to noise ratio followed by a simple classification algorithm. The "hard machine-learning" kind generalizes fairly well across sessions and across subjects, but requires a substantial amount of training data. Furthermore, it is often computationally intensive. The opposite happens for the "spatial filtering" kind, where bad generalization capabilities are compensated by a fast training and lower computational cost. In light of this situation it has been stated that "the field would benefit from a new paradigm in research development that focuses on robust algorithm development" (Obeid and Picone, 2013). It has also been recommended to start regarding the pre-processing, feature extraction and classification not as isolated processes, but a s a whole (Mak et al., 2011).

---

[1] http://bmiconference.org/wp-content/uploads/files/Future_BNCI_Roadmap.pdf

In this article we describe a classification paradigm possessing both good generalization and fast learning capabilities. The approach, based on recent advances in Riemannian geometry, supports a new BCI modes of operation. To proceed in this direction we have moved from the trends currently followed by the BCI community and by the specification of the characteristics a BCI should possess. Among recent trends in BCI research we find:

1. The conception, analysis and testing of *generic model classifiers*, allowing the so-called *transfer learning*, whereas data from other sessions and/or other subjects is used to increase the performance of low-performance users and to initialize a BCI so as to start using it without calibration (Colwell et *al*., 2013; Herweg, Kaufmann and Kübler, 2013; Kindermans and Schrauwen, 2013; Jin et al., 2012). In this direction is also relevant the use of unsupervised classifiers (Kindermans, Verstraeten and Schrauwen, 2012).

2. The conception, development and maintenance of world-wide massive *databases* (e.g., Obeid and Picone, 2013). Such a resource is necessary to boost research by allowing massive testing of algorithms. It also enables the systematic study of the source of variation in individual EEG patterns and their relation to BCI capabilities and individual attainable performances. Finally, it yields a smart initialization of a BCI (by specific transfer learning), which is necessary to use effectively a BCI without calibration.

3. The *continuous (on-line) adaptation of the classifier*, which combines the smart initialization mentioned above, in that the adaptation ensures that optimal performance is achieved regardless how good the initialization is. It also allows keeping optimal performance by adapting to mental and environmental changes during the session, achieving the sought reliability (Panicker, Puthusserypady and Sun, 2010; Schettini et *al*., 2013).

Other current lines of research that should be taken into consideration in designing a new generation of BCIs include:

I) The improvement of the performance by *dynamic stopping*, that is, minimizing the amount of data necessary to send a command (the duration of a BCI trial, e.g., the number of repetitions in the P300 speller) while keeping the same performance (Mainsah et *al*., 2013; Kindermans and Schrauwen, 2013; Schettini et *al*, 2013).

II) The improvement of the *BCI interface*, e.g., the introduction of *language models* in BCI spellers for letter and word prediction (Mainsah et *al*., 2013; Kaufmann et *al*., 2012; Kindermans and Schrauwen, 2013).

III) The improvement of the *BCI modality* itself, e.g., for P300-based BCI, the use of faces for flashing to improve the accuracy (Kaufmann et *al*., 2012), the use of random group or pseudo-random group flashing instead of row-column flashing (Congedo et *al*., 2011; Jin et *al*., 2011; Townsend et *al*., 2010), the use of inter-stimulus intervals randomly drawn from an exponential distribution and not constant (Congedo et *al*., 2011), etc. For SSVEP-based BCI improvements of the modality include the use of precise tagging of the flickering so as to use phase information (e.g., Jia et *al*., 2011), the use of random flickering sequences (code modulation: e.g., Bin et *al*., 2011), etc.

IV) *Multi-subject BCIs*, that is, BCI systems controlled by several users, in proximity one to the other or remotely connected (Bonnet et *al*., 2013; Schultze-Kraft et *al*., 2013). Besides allowing remote social operation, this functioning has potential to achieve 100% accuracy on single trials by combining the data of several users.

V) *BCI Hybridization*, that is, the combination of several BCI modalities on the same interface to increase the bit rate, the ergonomy, usability and the accuracy (e.g., Lee and Park, 2013).

A proposition for a BCI realizing characteristics 1-3, while keeping in mind points I-V, is presented schematically in fig. 1. See the caption of the figure for a generic description of such a BCI. In order to achieve such a functioning, we hold that the BCI processing and classification core should possess the following characteristics:

a) It should be *accurate* in general, as compared to state of the art approaches.
b) It should be *reliable*, that is, it should maintain as much constant as possible its functions and accuracy in routine circumstances, as well as in hostile or unexpected circumstances.
c) It should perform generally well as *initialized with generic parameter*, even for a naive user, that is, it should possess good generalization abilities both cross-subject and cross-session.
d) It should *learn fast* the individual characteristics and then maintains optimality, *adapting* fast to the mental state of the user and to environmental changes.
e) It should be *universal*, that is, applicable to all BCI paradigms (hence to hybrid systems).
f) It should be *algorithmically simple*, so as to be robust and usable in unsupervised on-line operation.
g) It should be *computationally efficient*, so as to work on small portable devices in line with the current trend in portability of micro-electronic devices.
h) It should be able to readily generalize to the *multi-user* settings

The BCI signal processing and classification framework we describe here possesses *all* characteristics a)-h). We show by means of real offline and online data that an effective "traditional" BCI can be obtained with a simple classification algorithm and very little pre-processing, regardless the chosen modality. Then we show that the framework is accurate, that it possesses good across-subject and across-session generalization properties, that it behaves well in a fully adaptive mode of operation and that it generalizes straightforwardly to the multi-user setting. For the first time, feature extraction and classification are regarded as a whole process common to all modalities. The Riemannian geometry framework here presented is still today largely unknown to colleagues involved in EEG analysis and BCI, yet, we predict that Riemannian geometry will play shortly a prominent role in BCI classification, forming the core methodology for upcoming BCIs.

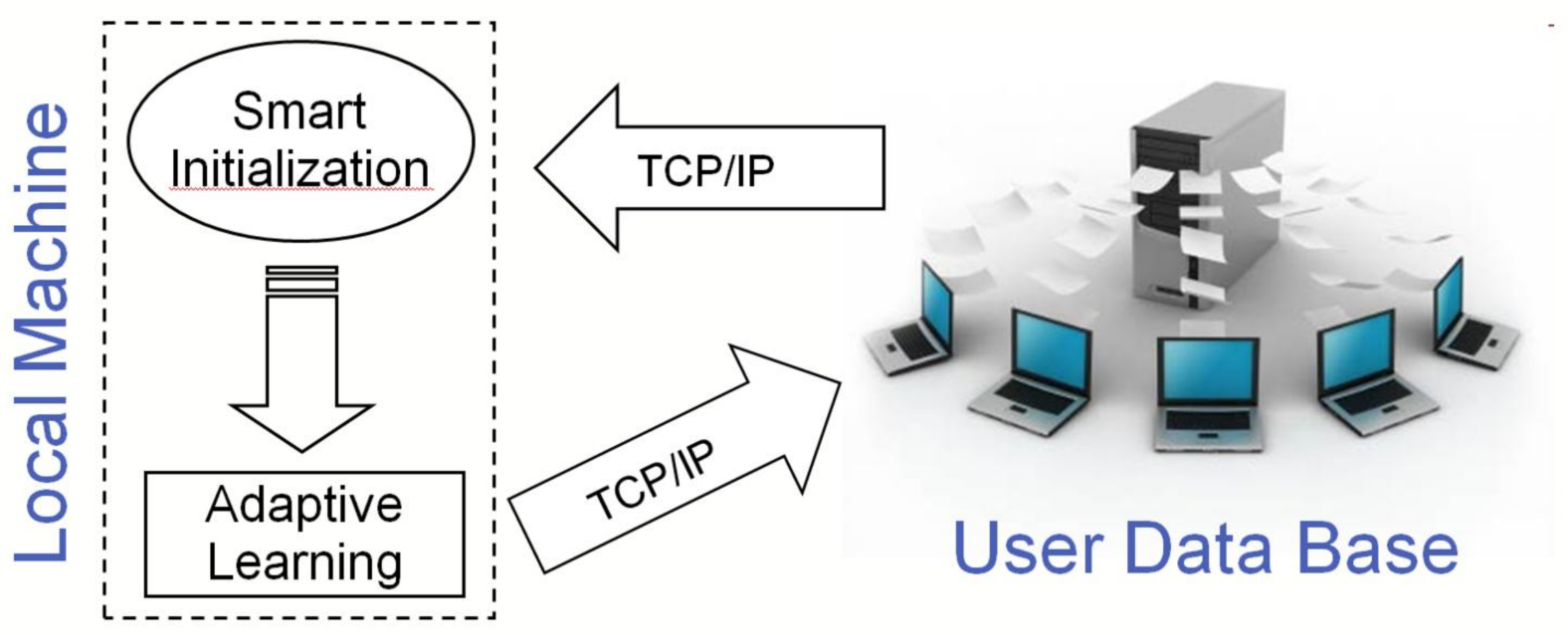

***Figure 1: Concept for a possible new generation of BCIs.*** *At start-up a BCI queries a database to obtain an initialization, possibly sending minimal EEG data of the user so that the database can elaborate a smart initialization that fits appropriately even a naive user. The BCI is operational straightaway, albeit suboptimal at the very beginning. While being used the BCI adapts to the user and send back the data to the database, along user information, so as to enrich the database and to allow smarter initializations in future sessions of the same user. Multiple subjects may use at the same time the same BCI, in which case the core of the BCI may be located on the server, so to exploit the multitude of data to increase performance.*

## Material and Methods

The task of a BCI is to classify single trials. First of all we specify a generic model for the observed data. Let $\boldsymbol{x}(t) \in \Re^N$ be the zero-mean EEG data vector for $N$ electrodes at discrete time sample $t$ and let $\boldsymbol{X}_z \in \Re^{N.T}$ be a trial (a finite time-interval realization) comprised of $T$ samples belonging to class $z \in \{1,...,Z\}$. The trial data is always assumed having zero mean (e.g., after usual band-pass filtering), therefore the *sample covariance matrix* of a given trial $\boldsymbol{X}_z$ in a wide-sense stationary (Yeredor, 2010) belonging to class $z$ is given by

$$\boldsymbol{C}_z = 1/(T-1)\left(\boldsymbol{X}_z \boldsymbol{X}_z^T\right). \tag{1}$$

Assuming a multivariate Gaussian distribution the *Wishart matrix* $\boldsymbol{\Sigma}_z$ is the only parameter of the data distribution, assumed unique for each class, which is indicated such as

$$\boldsymbol{x}_z(t) \sim \mathrm{N}\left(\boldsymbol{0}, \boldsymbol{\Sigma}_z\right) \tag{2}$$

and (1) is a sample covariance matrix estimation of the Wishart matrix. We shall describe a classification algorithm that can be applied *generically* with an appropriate definition of the "covariance matrix", so as to capture the relevant information of the trials. The relevant information, hence the band-pass filtering and the form of the covariance matrix, depends on the modality at hand, however we require the rest of the signal processing chain be the same for all paradigms. Note that in this article we employ the term "covariance matrix" referring to a generic structured symmetric positive definite matrix, which is in general different depending on the modality and of which the sample covariance matrix (1) typically is just a block. Hence, we will actually assume a multivariate Gaussian distribution as in (2) not for the observed data $\boldsymbol{x}(t)$, but for an *extended definition of observed data*, so as to allow well separated associated Wishart matrices and covariance matrices estimated on single-trials. If we succeed in this endeavor, the classification task becomes straightforward in the context of Riemannian geometry. Let us now precise this framework and comment on the requirements of good generalization and fast learning.

### *The Classification Framework*

It does not matter if we deal with motor imagery (MI) trials, steady-state evoked potentials (SSEP) trials, or event-related potentials (ERP) trials, in BCI operation we have a number of *training* (labeled) trials $\boldsymbol{X}_z$ for each class $z \in \{1,\ldots,Z\}$. The classification task consists in assigning an *unlabeled* trial $\boldsymbol{X}$, from which a special form of covariance matrix $\boldsymbol{C}$ is computed, to one of the $Z$ classes. Using training data we may compute a "mean" covariance matrix representing each one of the $Z$ classes, denoted $\boldsymbol{M}_1$ ,..., $\boldsymbol{M}_Z$, and then simply assign the unlabeled trials to the class which mean is the closest. In order to do so we need an appropriate *metric* to estimate the class *means* based on labeled trials and to assess the *distance* between the unlabeled trials and the means, denoted here such as $\delta_R\left(\boldsymbol{C} \leftrightarrow \boldsymbol{M}_z\right)$. This is what we obtain thanks to the Riemannian geometry framework. Our generic classification algorithm is summarized as it follows:

| *Universal MDM BCI Classifier (3)* |
| --- |
| *- Given a number of training trials* $\mathbf{X}_z$ *for each class* $z\in\{1,\dots,Z\}$ *do appropriate preprocessing, estimate an appropriate form of covariance matrix* $\mathbf{C}_z$ *and estimate their Z class means* $\mathbf{M}_1,\dots,\mathbf{M}_Z$. <br> *- For unknown trial* $\mathbf{X}$ *do the same preprocessing, estimate the same form of covariance matrix* $\mathbf{C}$ *and assign to class k as per* <br> $\underset{z}{argmin}\ \delta_R\left(\mathbf{C}\leftrightarrow\mathbf{M}_z\right)$, <br> *that is, to the class which mean is the closest to the covariance matrix, according to distance* $\delta_R$. |

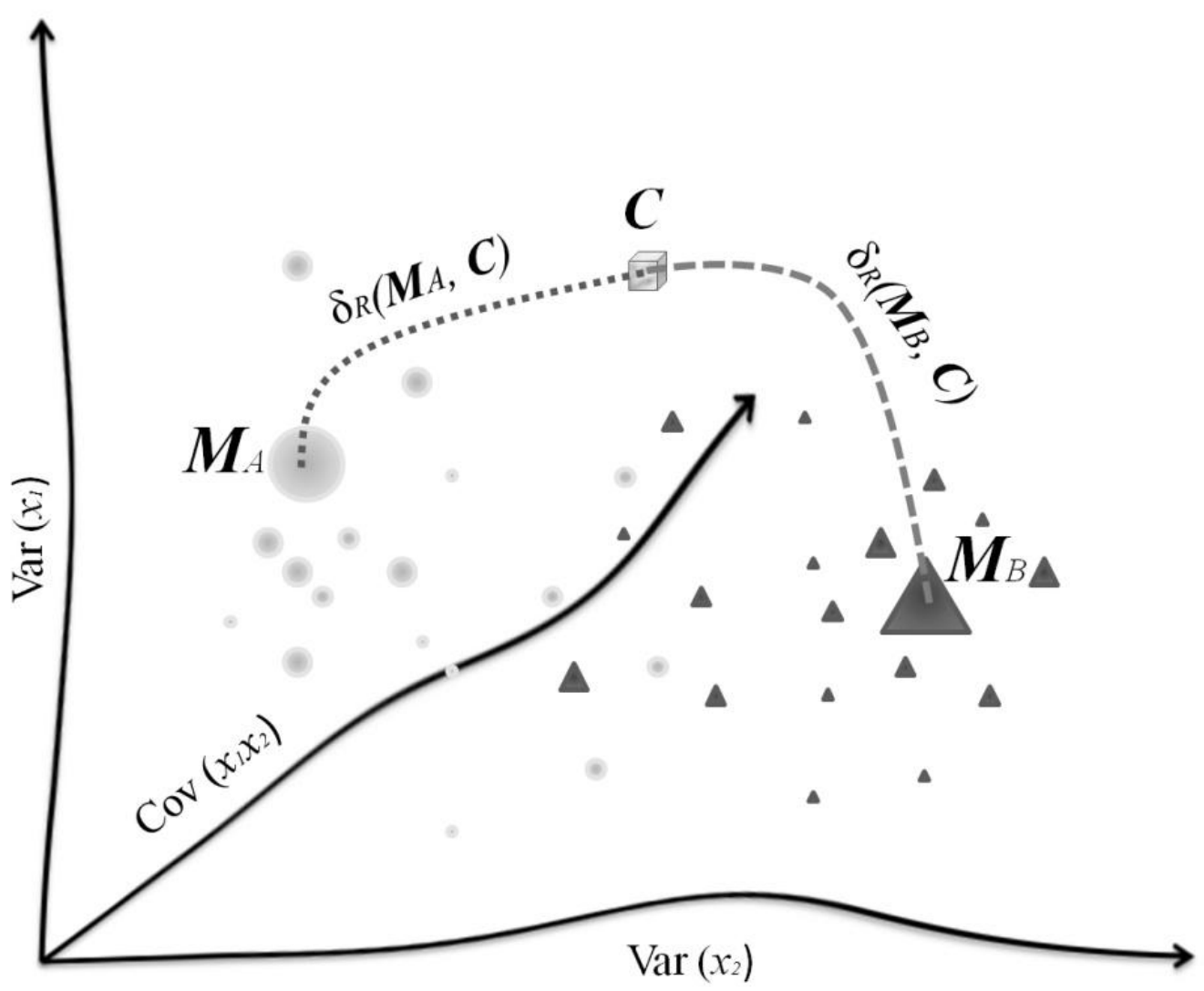

***Figure 2: The Minimum Distance to Mean (MDM) Classifier.*** *We illustrate the MDM algorithm with the example of 2x2 covariance matrices and two classes, labeled A and B. 2x2 covariance matrices have three non-redundant elements, which are the variance of the two process* $x_1(t)$ *and* $x_2(t)$ *and their covariance. We may then represent each covariance matrix as a point in 3D-space. The training (labeled) trials for class A are represented as spheres and for class B as pyramids. Given two geometric means* $\mathbf{M}_A$ *and* $\mathbf{M}_B$ *and un unlabeled trial with covariance matrix* $\mathbf{C}$ *(represented as a cube in the figure), the algorithm assigns the trial to the class which mean is the closest, according to an appropriate distance measure* $\delta_R$*. The distance measure is not linear, as can be appreciated in Appendix A. This is represented figuratively by curved lines in the figure. The MDM algorithm acts exactly in the same way for whatever dimension of the covariance matrices and whatever number of classes.*

This is the simplest classification method one can think of and is known as *minimum distance classifier* (MDM). The classification algorithm is illustrated in fig. 2 for the case of a two-class BCI ($Z$=2). It works exactly in the same way for whatever number of classes. As it is well known, defining the mean as the arithmetic mean and the distance as the Euclidean distance yields very poor classification accuracy (see for example Li et *al*., 2012). However, the message we want to convey here is that we do not need to complexify the classification algorithm or to apply sophisticated pre-processing and sophisticated spatial filtering or machine learning techniques. It turns out that an appropriate definition of the "covariance matrix", of the mean and of the distance performs as well as the most sophisticated methods that can be found in the literature. To make a metaphor, it appears that we have started a long time ago measuring distances with a biased ruler. Then we have developed complex instruments in order to replace the malfunctioning ruler. Finally, we have found a valid ruler, so that the complex instruments are no more necessary. Providing the valid ruler to measure distances is the main achievement of the Riemannian framework. The Riemannian geometry provides the natural framework to treat symmetric positive-definite matrices (SPD) and structured covariance matrices are of this type. Defining appropriate SPD matrices embedding relevant information depending on the data is our job, linking electrophysiological knowledge to mathematical formalism. We define an appropriate definition of *mean* and *distance* for SPD matrices in appendix A.

### *Smart Initialization (Cross-Subject and Cross-Session Generalization)*

We have said that a BCI processing chain should possess *both* fast adaptation abilities and good generalization across-session and across-subject. This is where the Riemannian framework proves advantageous as compared to the state of the art methods, which in general possess one, but not the other. The initialization of the classificator state by previous data, either coming from other individuals (cross-subject) or from previous sessions (cross-session) is sometimes named in the literature as *transfer learning*. Cross-subject transfer learning is the only option for a naive user. From the second use of the same BCI on we can use cross-session transfer learning as well. When an optimal subset of available data is used to specifically initialize the system for a given user we say the system is *smartly initialized*. For instance, how to optimally blend the cross-subject and cross-session initialization after the first session and what part of the database should be used to initialize the classifier for a given user are largely unexplored topics (see for example Schettini et *al*., 2013).

### *Adaptation*

Given an initialization we want to learn individual classification parameters so as to achieve optimal performance and adapt to environmental changes, mental state changes and any other intervening conditions that may affect the classification performance. We also want to do this as fast as possible. In order to do so we actually set up two parallel classification MDM algorithms, a *generic* one, based uniquely on transfer learning (i.e., a database), and an *individual* one, the latter being supervised or unsupervised. We consider here the supervised case. The classifier output will be given by a weighted sum of the two classifiers, say, with the two weights summing up to 1. In other words, the two classifiers are combined to produce a new classifier. The generic classifier will have weight 1 at the beginning of the session and smaller and smaller weight as the session progresses. The individual classifier will have weight zero at the beginning of the session for a naive user. For a user for which data from previous sessions is available the initial weight can be raised proportionally to the amount of training data. In any case the weight of the individual classifier will rise along the session and will approach 1 by the end of the session. How these two parallel classifiers should evolve over time, with or without supervision, is an intriguing and non-trivial research topic. Nonetheless, we take advantage of the MDM framework; indeed a MDM classifier is defined exhaustively by a set of mean covariance matrices and a distance metric. As a consequence, differently from other kinds of classifiers, several MDM classifiers can be combined easily by combining several distances, ensuring the stability and the consistency of the classification output.

### *Classification of Motor Imagery*

The sample covariance matrix as defined in (1) contains only *spatial information*. The diagonal elements hold the variance of the signal at each electrode and the off-diagonal elements hold the covariance between all electrode pairs. As such it suffices for classifying motor imagery (MI) trials because MI trials for different classes do indeed produce a different (scalp) spatial pattern, which is fully embedded in the structure of the sample covariance matrix (Pfurtscheller and Lopes da Silva, 1999). Then for MI we do not need to extend the definition of the data and we set simply

$$\boldsymbol{X}_z^{MI} = \boldsymbol{X}_z \, . \tag{4}$$

In case of MI-based BCIs there are as many classes (with associated training trials) as motor imagery tasks. A no motor imagery class can be added if sought. The only pre-processing step requested is filtering the data in the frequency band pass regions involved in motor imagery (e.g., 8-30 Hz). Then the MDM algorithm (3) applies using the regular form of covariance matrix as per (1). Extensive testing has proven that the MDM method is reliable, robust and accurate for motor imagery data (Barachant et *al*., 2010a, 2010b; 2011a, b; 2012a, b).

### *Classification of Event-Related Potentials (ERPs)*

For ERP-based BCI the standard sample covariance matrix is not efficient as the ERP features amplitude much smaller as compared to the background EEG, thus the spatial structure contained in the covariance matrix of a single trial does not hold sufficient information for classification. As a matter of fact the covariance matrix does not contain *any temporal information at all*, which is easily realized if we consider that shuffling at random the samples of trial $\boldsymbol{X}_z$ the sample covariance matrix (1) is unchanged. However ERPs have a specific time signature; it is this signature that differentiates an ERP from another or an ERP from the absence of the ERP, so this is the information we need to extract and embed in a "covariance matrix". In order to do so let us consider again a number of training trials $\boldsymbol{X}_z$, for $z\in\{1,\ldots,Z\}$ classes. In this case each class corresponds to a different ERP and a no-ERP class is usually added. For example, in P300-based BCI, one class is the *target* class, containing a P300, and the other is the *non-target* class ($Z$=2). Let us now construct the *super-trial*

$$\boldsymbol{X}_z^{ERP}=\begin{pmatrix}\overline{\boldsymbol{X}}_{(1)}\\ \cdots\\ \overline{\boldsymbol{X}}_{(Z)}\\ \boldsymbol{X}_z\end{pmatrix}\in\Re^{N(Z+1)\mathrm{x}T}, \tag{5}$$

where $\overline{\boldsymbol{X}}_{(1)}^T,\ \cdots,\overline{\boldsymbol{X}}_{(Z)}^T$ are the grand average ERPs obtained on the training data, previous sessions of the user or even on a database of other users (transfer learning) for each class; we call these grand-average ERPs the *temporal prototypes*. We specify a prototype for each class. Note that we have introduced index ($z$) in parenthesis to highlight the difference with the $z^{th}$ training class of the trial $\boldsymbol{X}_z$. Now, for a training trial $\boldsymbol{X}_z$ the covariance matrix of the super-trial has the following block structure:

$$\boldsymbol{C}_z=\frac{1}{(T-1)}\left(\boldsymbol{X}_z^{ERP}\left(\boldsymbol{X}_z^{ERP}\right)^T\right)=\frac{1}{(T-1)}\begin{pmatrix}\overline{\boldsymbol{X}}.\overline{\boldsymbol{X}}.^T & \left(\boldsymbol{X}_z\overline{\boldsymbol{X}}.^T\right)^T\\ \boldsymbol{X}_z\overline{\boldsymbol{X}}.^T & \boldsymbol{X}_z\boldsymbol{X}_z^T\end{pmatrix}\in\Re^{N(Z+1)\mathrm{x}\ N(Z+1)}, \tag{6}$$

$$\text{where }\overline{\boldsymbol{X}}.\overline{\boldsymbol{X}}.^T=\begin{pmatrix}\overline{\boldsymbol{X}}_{(1)}\overline{\boldsymbol{X}}_{(1)}^T & \cdots & \overline{\boldsymbol{X}}_{(1)}\overline{\boldsymbol{X}}_{(Z)}^T\\ \vdots & \ddots & \cdots\\ \overline{\boldsymbol{X}}_{(Z)}\overline{\boldsymbol{X}}_{(1)}^T & \cdots & \overline{\boldsymbol{X}}_{(Z)}\overline{\boldsymbol{X}}_{(Z)}^T\end{pmatrix}\in\Re^{NZxNZ} \tag{7}$$

$$\text{and }\ \boldsymbol{X}_z\overline{\boldsymbol{X}}.^T=\left(\boldsymbol{X}_z\overline{\boldsymbol{X}}_{(1)}^T,\cdots,\ \boldsymbol{X}_z\overline{\boldsymbol{X}}_{(Z)}^T\right)\in\Re^{NxNZ}. \tag{8}$$

Let us take a close look to the structure of this covariance matrix:

The $N$x$N$ diagonal blocks of $\overline{\boldsymbol{X}}.\overline{\boldsymbol{X}}.^T$ in (6) - see (7) for the relevant expansion - hold the covariance matrices of the $Z$ temporal prototypes and its $N$x$N$ off-diagonal blocks their cross covariance. All these blocks are not useful for classification since they, being based on fixed prototypes, do not change from trial to trial.

The $N$x$N$ block $\boldsymbol{X}_z\boldsymbol{X}_z^T$ in (6) holds the covariance matrix (1) of the trial $\boldsymbol{X}_z$, which contains the *spatial information* of the trial and will be little useful for classification, as we have said.

The $N$x$N$ blocks of $\boldsymbol{X}_z\overline{\boldsymbol{X}}.^T$ in (6) - see (8) for the relevant expansion - contains the cross-covariances between the trial and the $Z$ prototypes, that is, these blocks contain the *temporal covariances*. Notice that shuffling the samples of the trial now *does disrupt* these covariances. These blocks contain the relevant information for classification as the cross-covariance will be large only in the blocks where the class of the trial coincides with the class of the prototype. The means of the "super" covariance matrices $\boldsymbol{C}_z$ constructed as per (6) on training data, which we denote by $\boldsymbol{M}_1,\cdots,\boldsymbol{M}_z$ for the $Z$ classes, have of course the same structure as $\boldsymbol{C}_z$. With an unlabeled trial $\boldsymbol{X}$ we then construct the super-trial as per (5) and the corresponding covariance matrix $\boldsymbol{C}$ as per (6), where in both $\boldsymbol{X}$ replaces $\boldsymbol{X}_z$. Then, the classification is obtained as before using MDM (3). The only pre-processing required is to filter the data in the frequency band pass region containing the ERPs, typically 1-16 Hz. The exact choice of the band-pass region is not crucial for ERP classification. Extensive testing in our laboratory has shown that the method is reliable and robust, generalizes better than state of the art methods both across-session and across-subject and is prone to fast adaptation. We report in the result section new results corroborating this conclusion.

It is worth mentioning that often we deal only with the presence and absence of an ERP, as it is the case of P300-bases BCIs, where there are only two classes, a *target* (P300 present) and *non-target* (P300 non-present) class. In this case one can use a simplified version of super-trial (5) given by

$$\boldsymbol{X}_z^{P300}=\begin{pmatrix}\overline{\boldsymbol{X}}_{(+)}\\ \boldsymbol{X}_z\end{pmatrix}\in\Re^{2N\mathrm{x}T}, \tag{9}$$

where $\overline{\boldsymbol{X}}_{(+)}^T$ is the temporal prototype of the P300 (target class) and $z\in\{+,-\}$, with "+" indicating the target class and "– " indicating the non-target class. For a training trial $\boldsymbol{X}_z$ the covariance matrix of the super-trial has now the simpler block structure:

$$\boldsymbol{C}_z=\frac{1}{(T-1)}\left[\boldsymbol{X}_z^{P300}\left(\boldsymbol{X}_z^{P300}\right)^T\right]=\frac{1}{(T-1)}\begin{pmatrix}\overline{\boldsymbol{X}}_{(+)}\overline{\boldsymbol{X}}_{(+)}^T & \overline{\boldsymbol{X}}_{(+)}\boldsymbol{X}_z^T\\ \boldsymbol{X}_z\overline{\boldsymbol{X}}_{(+)}^T & \boldsymbol{X}_z\boldsymbol{X}_z^T\end{pmatrix}\in\Re^{2Nx2N}. \tag{10}$$

As in (6) the covariance of the prototype $\overline{\boldsymbol{X}}_{(+)}\overline{\boldsymbol{X}}_{(+)}^T$ does not change from trial to trial and is useless for classification.

The covariance of the trial $\boldsymbol{X}_z\boldsymbol{X}_z^T$ will be little useful for classification, as we have seen.

The temporal covariance between the prototype and the trial $\bar{\boldsymbol{X}}_{(+)}\boldsymbol{X}_z^T$ will be large if the trial pertains to a target and small if it does not, so (10) suffices to classify efficiently target and non-target trials, as we will show.

Equation (10) is the super-trial we have been using in Barachant and Congedo (2014) and Barachant et al. (2013) and we have found it equivalent to the more general form (5) for a two-class P300-based BCI. This is also the super-trial we have used for the ERP-based BCI results presented here below.

Finally, notice that in P300-based BCIs we classify after several P300 trials, that is, after several repetitions of exhaustive flashing of all elements. The classification then is based either on the cumulating sum of distances across repetitions or on a single distance obtained on the average trial computed across repetitions, the two approaches being equivalent.

### *Classification of Steady-State Evoked Potentials*

The same MDM method can be used for *SSEP* data classification as well. We make here the example of steady-state visually evoked potentials (SSVEP). The *Z* classes here represent *F* different flickering frequencies and a no-flickering class can be added as well if sought. In this case the relevant information is the diversity of the frequencies engendering oscillations in the visual cortex, while the spatial pattern may be the same for different frequencies. In order to exploit the frequency diversity we construct super trial

$$\boldsymbol{X}_z^{SSEP} = \begin{pmatrix} \boldsymbol{X}_{(1)} \\ \cdots \\ \boldsymbol{X}_{(F)} \end{pmatrix} \in \Re^{NF\text{x}T} \qquad (11)$$

where $\boldsymbol{X}_{(f)}^T$ is the trial filtered in the band-pass region for flickering frequency $f \in \{1,\ldots,F\}$. More simply, one may use the Fourier cospectra (Bloomfield, 2000) for the exact flickering frequencies. The covariance matrix of super-trial (11) has the following block structure:

$$\boldsymbol{C}_z = \frac{1}{(T-1)}\left[\boldsymbol{X}_z^{SSEP}\left(\boldsymbol{X}_z^{SSEP}\right)^T\right] = \frac{1}{(T-1)}\begin{pmatrix} \boldsymbol{X}_{(1)}\boldsymbol{X}_{(1)}^T & \cdots & \boldsymbol{X}_{(1)}\boldsymbol{X}_{(F)}^T \\ \vdots & \ddots & \vdots \\ \boldsymbol{X}_{(F)}\boldsymbol{X}_{(1)}^T & \cdots & \boldsymbol{X}_{(F)}\boldsymbol{X}_{(F)}^T \end{pmatrix} \in \Re^{NF\text{x}NF}. \qquad (12)$$

The *N*x*N* diagonal blocks holds the covariance matrices of the *F* frequencies. When comparing an unlabeled trial with the mean of the different classes, only the mean with the block indexing the frequency corresponding to the frequency of the trial will have large values. Thus the diagonal blocks will be useful for classification. On the other hand the off-diagonal blocks hold the cross-covariance between frequencies, thus are not very meaningful. We can put them to zero since the resulting matrix

$$\boldsymbol{C}_z = \frac{1}{(T-1)} \begin{pmatrix} \boldsymbol{X}_{(1)}\boldsymbol{X}_{(1)}^T & \cdots & \boldsymbol{0} \\ \vdots & \ddots & \vdots \\ \boldsymbol{0} & \cdots & \boldsymbol{X}_{(F)}\boldsymbol{X}_{(F)}^T \end{pmatrix} \in \Re^{NF\mathrm{x}NF} \quad (13)$$

is still symmetric positive definite. Given training data we estimate the class means $\boldsymbol{M}_1,\cdots,\boldsymbol{M}_z$. For an unlabeled trail $\boldsymbol{X}$ we compute the super trail with (11), where $\boldsymbol{X}$ replaces $\boldsymbol{X}_z$, then covariance matrix $\boldsymbol{C}$ using (13). Finally, we use again algorithm (3) to assign the unlabeled trial. The only pre-processing required is to filter the data in the frequencies corresponding to the SSVEP flickering frequencies or, equivalently, estimating the Fourier cospectra at the $F$ flickering frequencies. Note that if the phase of the SSVEP is known thanks to precise data tagging, as it is done in Jia et *al*. (2011), or code modulation is used (Bin et *al*., 2011), one can exploit both the frequential and the temporal information, constructing a super trial mixing the strategy used here for ERP (5) and for SSVEP (11).

### *Classification in the Multi-User Setting*

Finally, let us see how the proposed framework readily extend to the multi-user case. In multi-user BCIs multiple users interact at the same time with the same interface (Bonnet, Lotte and Lécuyer; Yuan et *al*., 2013; Schultze-Kraft et *al*., 2013). The interesting point of such set up is that we can in theory obtain a BCI with 100% accuracy on single-trial, given a sufficient number of subjects; instead of averaging data or classification scores across trials, we can do the same across subjects on a single-trial·. We can actually go a little further and exploit not only the multitude of trials, but also their expected synchronization between subjects. Let us make the example of P300-based BCIs. For other modalities the development is similar. As we have seen, for single-subject using a P300-based BCI the super-trials have form (9). For the multi-user (MU) case the trial for class $z\in\{+, -\}$, where + is the target class and – the non-target class, is given by

$$\boldsymbol{X}_z^{MU\ P300} = \begin{pmatrix} \bar{\boldsymbol{X}}_{(+)}^T \\ \boldsymbol{X}_{1z}^T \\ \cdots \\ \boldsymbol{X}_{Mz}^T \end{pmatrix} \in \Re^{(N(M+1)N}, \quad (14)$$

where $M$ is the number of subjects and $N$ the number of EEG sensors. Notice that the temporal prototype is still just one, as for the single-user case in (9), as it applies equally well to all users. The covariance matrix of (14) for the example case $M$=2 has form

$$C_z = \frac{1}{(T-1)}\left[ X_z^{MU\ P300}\left( X_z^{MU\ P300}\right)^T \right] = \frac{1}{(T-1)}\begin{pmatrix} \bar{X}_{(+)}\bar{X}_{(+)}^T & \bar{X}_{(+)}X_{1z}^T & \bar{X}_{(+)}X_{2z}^T \\ X_{1z}\bar{X}_{(+)}^T & X_{1z}X_{1z}^T & X_{1z}X_{2z}^T \\ X_{2z}\bar{X}_{(+)}^T & X_{2z}X_{1z}^T & X_{2z}X_{2z}^T \end{pmatrix}. \quad (15)$$

Block $\bar{X}_{(+)}\bar{X}_{(+)}^T \in \Re^{NxN}$ is again the covariance matrix of the temporal prototype. This does not change across trials and has no value for classification.

The off-diagonal blocks $X_{1z}\bar{X}_{(+)}^T$ and $X_{2z}\bar{X}_{(+)}^T \in \Re^{NxN}$ (or their transpose $\bar{X}_{(+)}X_{1z}^T$ and $\bar{X}_{(+)}X_{2z}^T$) hold the covariances between the trial of the subjects and the prototype and are relevant for classification just as in (6); the only difference is that instead of having only one of such covariance, now we have two of them, increasing accordingly the classification power.

Moreover, consider now off-diagonal block $X_{2z}X_{1z}^T$. This (or its transpose $X_{1z}X_{2z}^T$) holds the *covariance between the trial of the two subjects*; since the P300 response is synchronized, this covariance will be large for the target trials only, thus it hold some information useful for classification.

Notice that classifying based on (15) does not amount to simply add or combine the data of each individual. This is what we would obtain putting to zero the off-diagonal blocks $X_{2z}X_{1z}^T$ and $X_{1z}X_{2z}^T$ of (15). Here we are actually exploiting also the synchronization of the ERP response and the covariance of the synchronized response of the two subjects. The same goes with more than two users, wherein all pair-wise cross-subject covariances may be exploited. Therefore, the present Riamannian framework is a true multiple-user approach like joint blind source separation (Anderson at al., 2012; Congedo et *al*., 2012; Li et *al*., 2009; Vía et al., 2011).

## Results and Discussion

### *Motor imagery BCI data*

We have applied the MDM method to dataset 2a of BCI Competition IV (2008), provided by the Institute for Knowledge Discovery (Laboratory of Brain-Computer Interfaces), Graz University of Technology. The data set includes nine subjects involved in a four-class (*Z*=4) two-session motor imagery-based BCI experiment. The four classes were right hand, left hand, feet, tongue. EEG data was acquired by means of 22 electrodes concentrated on and around the sensorymotor areas. The trials were band-pass filtered in the range 8-30 Hz. Two seconds of data in each trial were used for the analysis. We consider both binary classification of each class against the others - this is what the state-of-the-art common spatial pattern (CSP) does better (Lotte et *al*., 2007) – and the true multiclass case, where the four classes are treated altogether. The MDM handles equally well and in the same way both the binary and the multiclass case.

We present results of the offline analysis to compare MDM to state of the art competitors. For binary classification we compare the MDM against CSP + LDA (linear discriminant analysis) classification algorithm (Lotte and Guan, 2011)[2]. Three pairs of CSP filters were retained. This is the unique parameter to be set with this approach. For multiclass classification we compare the MDM with the BSS approach proposed by Grosse-Wentrup and Buss (2008). Their method consists in the approximate joint diagonalization (AJD) of the four class covariance matrices, selection of the eight best filters using a mutual information criterion and a sparse logistic regression (LR) classifier[3]. For this approach also, the number of filters must be set. On the other hand, the MDM approach is fully automatic. The results concern the *cross-session performance*, that is, the algorithms are trained on one session and tested on the other. This is a more difficult test-bed as compared to cross-validation within the same session.

Results in term of accuracy (percent of correctly classified trials) for all subjects, the two sessions, one against the other, and for all methods are given in table 1. The chance level is 25%. We compared statistically the average performance in term of percent correct classified trial (accuracy) of the MDM and the state of the art competitors. For the multiclass classification the MDM proved marginally superior on the average of the 18 sessions as compared to the BSS + LR method (paired $\text{t-test}_{(17)}=1.9$, $p=0.074$, two-tailed). For the average of all the binary classification there was no difference between the MDM and the CSP+LDA method.

In Gouy-Pailler et *al*. (2010) we have developed a BSS (blind source separation) method for motor imagery classification exploiting the non-stationarity of the ERD/ERS during the trial. The method proposed by Grosse-Wentrup and Buss (2008) was the starting point of our investigation. The non-stationarity extension developed in Gouy-Pailler et *al*. (2010) is obtained estimating several covariance matrices in successive time intervals within the trial and diagonalizing all these matrices for all classes simultaneously. It was shown to be superior to the method of Grosse-Wentrup and Buss (2008). The mean (sd) obtained in Gouy-Pailler et *al*. (2010) for the cross-session accuracy was 63.8 (12.28). This mean is directly comparable to the means reported in table 1 and is not significantly superior to the MDM multiclass. We conclude that the MDM for MI data performs as well as the most sophisticated method found in the literature, but is fully automatic (no parameter to be set) and respect the requirements of a next generation of BCI.

---

[2] *For the CSP + LDA we used the code of Dr. Lotte, available under request by e-mail.*

[3] *For the BSS + LG we have used the code available at* http://people.kyb.tuebingen.mpg.de/moritzgw/MulticlassCSP.zip.

To take a closer look at the results we plotted the accuracy of the MDM against its competitor (fig. 3). It appears then the result we have observed consistently when comparing the Riemann MDM approach with state of the art approaches: the performance of the MDM approach is more or less equivalent for subjects performing well, while it is better for subjects performing poorly (see the position of the dots in the lower-bottom corner of the plots). This behavior springs from the robustness of the Riemannian metric (Appendix A).

***Table 1: Accuracy for MI data.*** *Results for each subject (M=9) using as training data session 1 and test data session 2 (rows "Session 1->Session 2") and vice versa (rows "Session 2->Session 1"), for the binary classification of all pairs of classes (numbered 1 to 4), the average of all binary classifications (columns "Ave") and for the multiclass classification (columns "4-class"). The MDM method applies to both the binary and multiclass classification. As state of the art competitors we used the CSP+LDA for binary classification and the BSS+LR for multiclass classification.*

| | ***MDM*** | | | | | | | | ***CSP + LDA*** | | | | | | | ***BSS + LR*** | |
|---|---|---|---|---|---|---|---|---|---|---|---|---|---|---|---|---|---|
| *Subject* | 1 / 2 | 1 / 3 | 1 / 4 | 2 / 3 | 2 / 4 | 3 / 4 | **Ave** | ***4-class*** | 1 / 2 | 1 / 3 | 1 / 4 | 2 / 3 | 2 / 4 | 3 / 4 | **Ave** | ***4-class*** | |
| *1* | 93.75 | 96.53 | 98.61 | 97.22 | 99.31 | 70.14 | **92.59** | ***78.82*** | 93.06 | 98.61 | 98.61 | 97.22 | 100.00 | 69.44 | **92.82** | ***76.74*** | Session 1 -> Session 2 |
| *2* | 63.19 | 78.47 | 68.06 | 74.31 | 72.22 | 74.31 | **71.76** | ***46.88*** | 50.69 | 68.75 | 67.36 | 81.25 | 63.89 | 69.44 | **66.90** | ***43.40*** | |
| *3* | 94.44 | 89.58 | 86.81 | 95.14 | 97.92 | 66.67 | **88.43** | ***70.83*** | 96.53 | 94.44 | 94.44 | 93.06 | 96.53 | 69.44 | **90.74** | ***76.04*** | |
| *4* | 75.00 | 88.89 | 88.19 | 92.36 | 84.72 | 62.50 | **81.94** | ***61.81*** | 70.14 | 78.47 | 86.81 | 88.89 | 85.42 | 56.94 | **77.78** | ***55.21*** | |
| *5* | 63.19 | 73.61 | 72.22 | 72.22 | 76.39 | 70.14 | **71.30** | ***50.00*** | 59.03 | 63.19 | 68.75 | 68.75 | 65.28 | 70.83 | **65.97** | ***35.07*** | |
| *6* | 71.53 | 71.53 | 64.58 | 65.28 | 63.89 | 72.92 | **68.29** | ***47.57*** | 68.06 | 59.03 | 71.53 | 63.19 | 65.97 | 67.36 | **65.86** | ***44.44*** | |
| *7* | 72.92 | 91.67 | 88.19 | 90.97 | 85.42 | 78.47 | **84.61** | ***66.32*** | 79.86 | 97.92 | 95.14 | 99.31 | 97.22 | 81.25 | **91.78** | ***63.19*** | |
| *8* | 96.53 | 85.42 | 82.64 | 90.28 | 76.39 | 70.83 | **83.68** | ***72.57*** | 93.75 | 87.50 | 90.97 | 86.81 | 91.67 | 82.64 | **88.89** | ***69.44*** | |
| *9* | 91.67 | 93.06 | 97.22 | 72.22 | 81.94 | 90.28 | **87.73** | ***74.31*** | 92.36 | 95.14 | 95.14 | 84.72 | 81.94 | 88.89 | **89.70** | ***79.17*** | |
| *1* | 74.31 | 94.44 | 95.14 | 92.36 | 98.61 | 79.86 | **89.12** | **71.18** | 77.78 | 95.14 | 95.83 | 91.67 | 99.31 | 80.56 | **90.05** | **73.61** | Session 2 -> Session 1 |
| *2* | 50.00 | 76.39 | 53.47 | 77.08 | 74.31 | 77.78 | **68.17** | **50.00** | 50.00 | 74.31 | 59.03 | 59.72 | 60.42 | 77.78 | **63.54** | **29.51** | |
| *3* | 88.89 | 82.64 | 92.36 | 85.42 | 94.44 | 81.25 | **87.50** | **74.65** | 91.67 | 86.81 | 96.53 | 90.97 | 93.75 | 85.42 | **90.86** | **78.47** | |
| *4* | 65.28 | 72.22 | 78.47 | 70.83 | 72.92 | 68.75 | **71.41** | **49.65** | 66.67 | 72.92 | 81.25 | 79.17 | 71.53 | 71.53 | **73.84** | **42.01** | |
| *5* | 63.89 | 65.28 | 72.22 | 66.67 | 72.22 | 59.03 | **66.55** | **37.85** | 61.11 | 51.39 | 70.83 | 54.17 | 67.36 | 53.47 | **59.72** | **26.39** | |
| *6* | 61.81 | 70.14 | 59.72 | 68.06 | 61.11 | 63.19 | **64.00** | **42.71** | 70.83 | 76.39 | 61.11 | 66.67 | 65.97 | 68.06 | **68.17** | **38.19** | |
| *7* | 73.61 | 87.50 | 85.42 | 87.50 | 87.50 | 77.78 | **83.22** | **65.97** | 66.67 | 91.67 | 95.83 | 98.61 | 97.22 | 86.11 | **89.35** | **67.71** | |
| *8* | 94.44 | 79.17 | 87.50 | 88.89 | 87.50 | 79.86 | **86.23** | **71.53** | 97.92 | 83.33 | 96.53 | 93.06 | 92.36 | 86.11 | **91.55** | **75.35** | |
| *9* | 81.25 | 88.19 | 98.61 | 83.33 | 89.58 | 88.89 | **88.31** | **72.92** | 91.67 | 93.75 | 100.00 | 81.25 | 92.36 | 84.03 | **90.51** | **74.65** | |
| *mean* | ***76.43*** | ***82.48*** | ***81.64*** | ***81.67*** | ***82.02*** | ***74.04*** | ***79.71*** | ***61.42*** | ***76.54*** | ***81.60*** | ***84.76*** | ***82.14*** | ***82.68*** | ***74.96*** | ***80.45*** | ***58.26*** | |
| *sd* | 14.06 | 9.43 | 13.75 | 10.75 | 11.58 | 8.58 | 9.44 | 13.15 | 16.09 | 14.33 | 14.27 | 14.03 | 14.67 | 10.24 | 12.30 | 18.80 | |

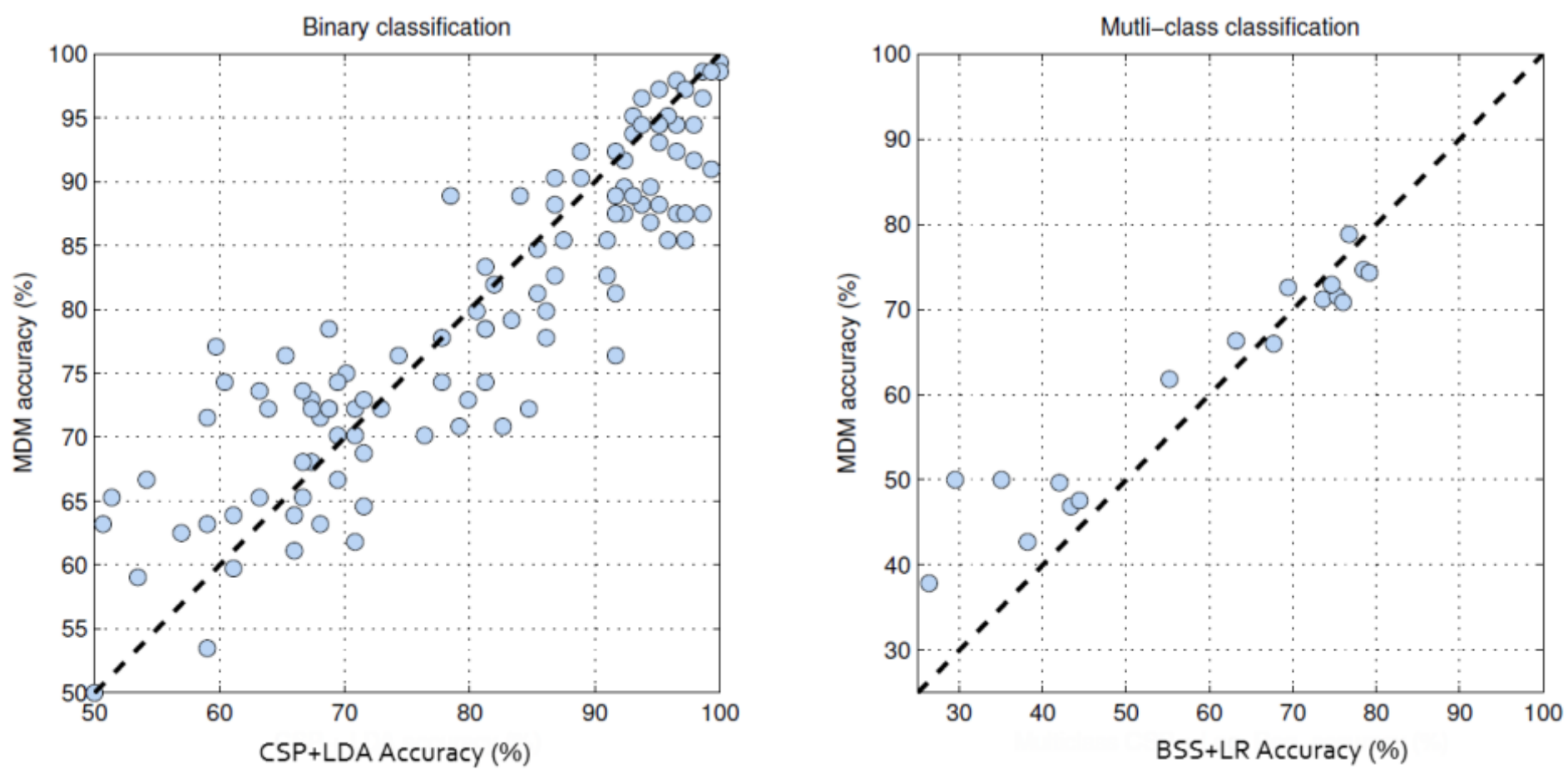

***Figure 3: Motor Imagery Performance.*** *Performance in terms of percent correctly classified trials (accuracy), for the binary classification and for the multiclass classification. Each dot represents a subject and a session and has coordinates given by the accuracy of the MDM method against the state of the art competitor.*

## *P300 BCI data*

We present several results issued from an extensive experiment performed in our laboratory in Grenoble. 24 subjects performed one session of the P300-based BCI video-game *Brain Invaders* (Congedo et *al*., 2011). Seven of these subjects performed seven more sessions, twice a week, for a total of eight sessions. Each session consisted of two runs of the Brain Invaders, one using the typical training-test procedure (non-adaptive mode) and the other without any training using an initialization and an adaptation scheme (adaptive mode), as presented in the method section. The two runs looked exactly identical to the subjects, in that in both cases a training session preceded a test session. However, the BCI classification was different in the two runs. For the non-adaptive mode the universal MDM classification algorithm (3) was trained in the training run and applied in the test run. For the adaptive mode the classification was initialized using a database for the first session and then with the cumulated data of previous sessions of the user, so that training data was simply discarded for the adaptive mode. The order of the two runs was randomized and the design was double-blinded; at any time neither the subject nor the experimenter could know in what mode the BCI was running. For the MDM algorithm we use as definition of the trial (9) and the corresponding covariance matrices obtained by equation (10).

Data was acquired with a g.USBamp amplifier (g.Tec, Graz, Austria) using 16 active wet electrodes positioned at Fp1, Fp2, Afz, F5, F6, T7, Cz, T8, P7, P3, Pz, P4, P8, O1, Oz, O2, referenced at the right earlobe with a cephalic ground and sampled at 512 Hz. In online operation and for offline analysis EEG data were band-pass filtered in the range 1-16 Hz and downsampled to 128 Hz.

We present both online results and offline results, the latter in order to compare the MDM algorithm with two popular state of the art algorithms: XDAWN (Rivet et *al*., 2011) and the stepwise linear discriminant analysis (SWLDA, Farwell and Donchin, 1988). For XDAWN the two most discriminant spatial filters were retained. EEG data was then spatially filtered, decimated to 32 Hz and vectorized so as to classify the obtained 32x2 features with a regularized linear discriminant analysis (LDA), using an automatic setting of the regularization parameter (Ledoit and Wolf, 2004; Vidaurre at *al*., 2009). For the SWLDA, EEG data were decimated to 32 Hz and vectorized so as to feed the classifier with the obtained 32x16 features.

We begin by presenting several offline results of the performance pertaining to the non-adaptive mode, including the classic training-test setting and the cross-subject and cross-session initialization comparing several classifiers. We also present the online results obtained in the adaptive and non-adaptive mode. These latter results are the most relevant as they report the actual performance achieved by the universal MDM algorithm in real operation. All performance results for this experiment are reported in terms of AUC (area under the curve). Before we detail the performance results, let us visualize the structure of the covariance matrix (10) for one subject. This illustrates well the rationale behind the choice of this form of the covariance matrix for ERP-based BCI. See Fig. 4 and its caption for details.

*Offline results: the "classic" training-test mode.*

Fig. 5 shows the grand average (7 subjects x 8 sessions) AUC accuracy criterion for the three classification methods, obtained training the classifiers on the training run and testing on the test run ("Classic" column). Table 2 reports the detailed results for each subject and session. Paired t-tests revealed that the mean AUC obtained by the MDM is significantly superior to the mean AUC obtained by the SWLDA method ($t_{(55)}$= 3.377, $p$=0.001), and equivalent to the mean AUC obtained by XDAWN.

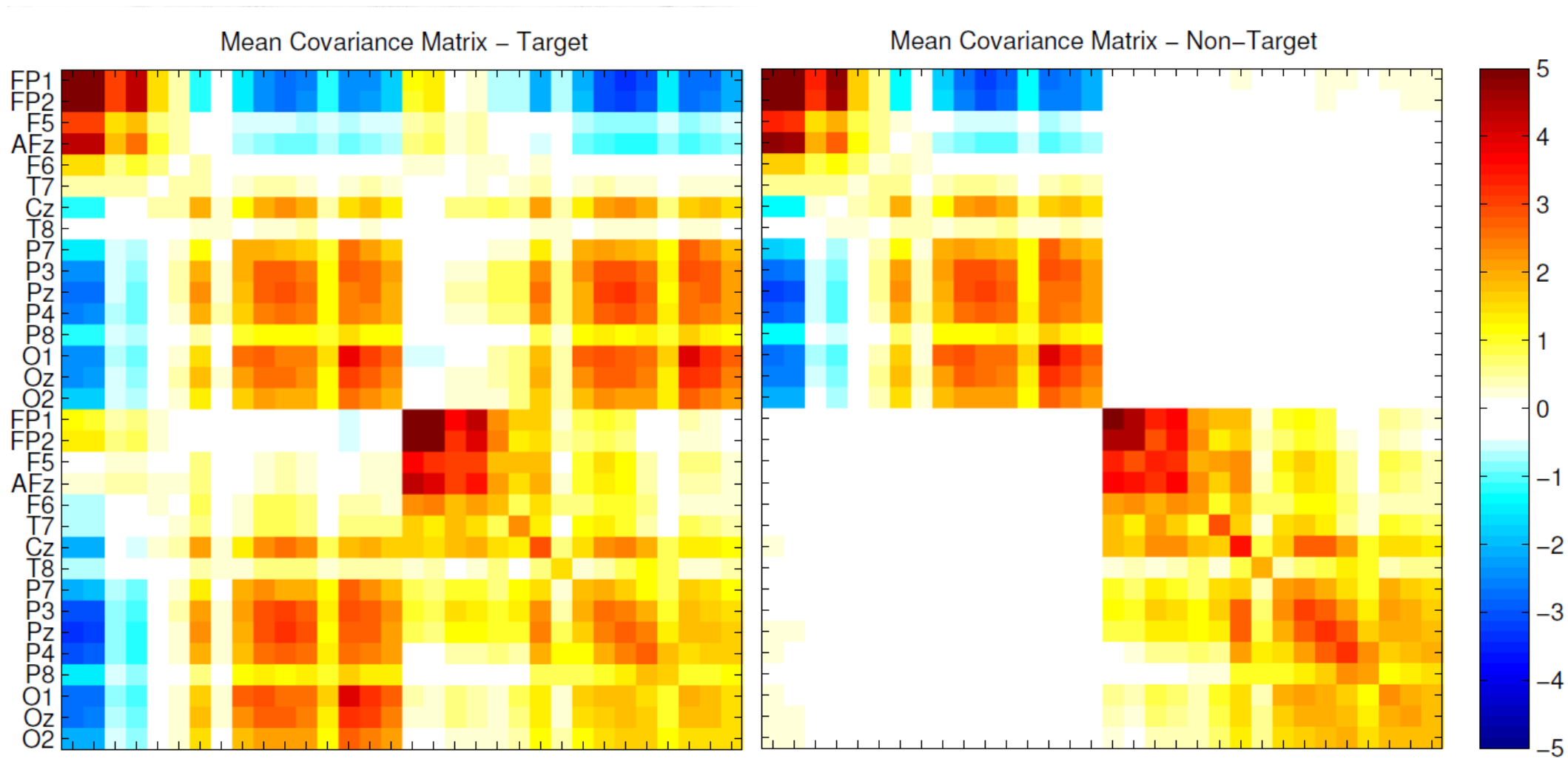

***Figure 4: Image of the Covariance matrix with form (10).*** *The covariance matrix is computed on one subject using the Brain Invaders P300-based BCI. The matrix is divided in four 16x16 blocks. The upper-left diagonal block is the grand-average sample covariance matrix of the target prototype computed on the other six subjects (cross-subject initialization). This block is the same on the left (target) and right (non-target) part of the figure. The lower-right diagonal block is the sample covariance matrix of the average ERPs obtained on the subject with 106 flashes for target (left) and 530 for non-target (right). The off-diagonal blocks are the temporal covariances between the prototype and the average ERPs; this covariance is high only for the target class since only in this case the signal produced by target flashes temporally correlates with the prototype. All covariance matrices are computed on the ERPs recorded 1s after the flash. The diagonal blocks are scaled so as to make the plot readable.*

*Offline results: the cross-subject initialization.*

These results are obtained using a leave-one-out method. Fig. 5 shows the grand average (7 subjects x 8 sessions) AUC accuracy criterion for the three classification methods obtained training the classifiers on the test data of all subjects excluding the one on which the performance are computed ("Cross-subject" column). Table 2 reports the detailed cross-subject results for each subject and session. As compared to the classic mode the average AUC with cross-subject transfer learning is significantly lower for all classification methods ($p<0.002$ for all of them). This is an expected result as no information at all about the subject actually using the BCI is provided to the classifiers. Paired t-tests comparing the average performance of the three classification methods in the cross-subject mode reveal that the average AUC obtained by the MDM is marginally superior to the average AUC obtained by the SWLDA ($t_{(55)}= 1.676$, $p=0.099$) and by XDAWN ($t_{(55)}= 1.755$, $p=0.085$).

*Offline results: the cross-session initialization.*

These results are also shown in Fig. 5 ("Cross-session" column). The mean AUC is obtained initializing the classifier with any possible combination of $S$ number of sessions among the eight available sessions and testing on the remaining 8-*S* sessions. The results are given for *S* in the range 1,…,7 and correspond to the average of all subjects and all combinations (which number depends on *S*). The MDM algorithm proves superior both in the rapidity of learning from previous subject's data and in the performance attained for all values of *S*. Note that XDAWN, which is a spatial filter approach, performs fairly well even when only one session is available for training, but its performance grows slowly as more data is available for training. This is because the spatial filter is influenced negatively by the difference in electrode placements across sessions and, in general, by all factors that may change from one session to the other. On the other hand the SWLDA classifier performs poorly when only one session is available for training, however it learn fast as the number of available sessions increase. This is because the SWLDA, being an "hard machine learning" approach, tends to perform well only when a lot of training data is available. So, XDAWN possesses fast learning capabilities, but lacks good transfer learning, whilst the opposite holds for SWLDA. The MDM algorithm possesses both desirable properties.

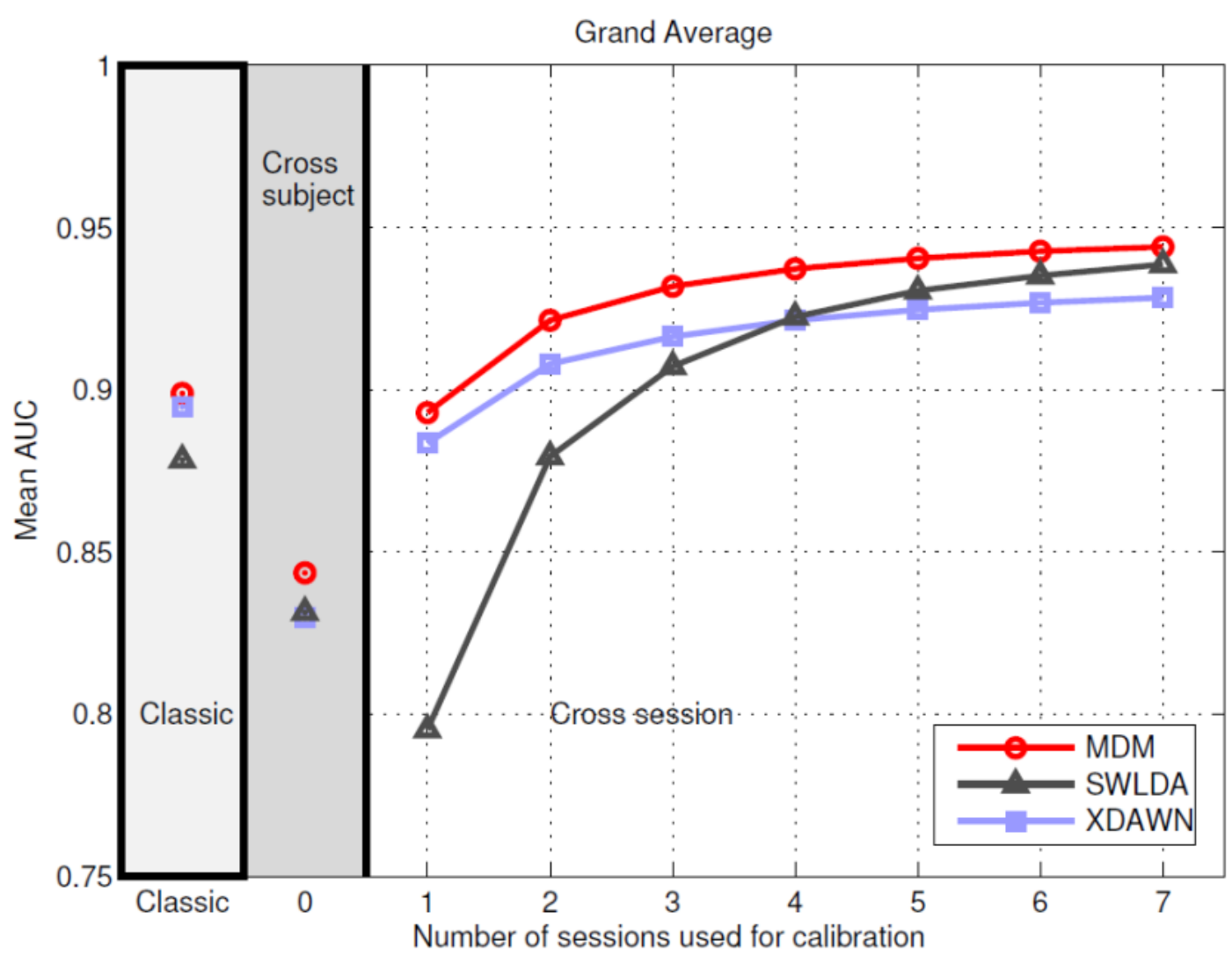

***Figure 5: Classic (training-test), cross-subject and cross-session offline AUC performance for the P300-based Brain Invaders BCI experiment.*** *Results are the grand average of 7 subjects playing 8 sessions of the Brain Invaders. See text for details.*

***Table 2: Classic (training-test) and cross-subject AUC performance for the P300-based BCI experiment.*** *AUC performance is given separately for each one of the seven subjects (SS) and of the eight sessions (Sess) of the Brain Invaders. See text for details.*

| | | Classic | | | Cross-Subject | | | | Classic | | | Cross-Subject | | |
|---|---|---|---|---|---|---|---|---|---|---|---|---|---|---|
| **Sess** | **SS** | **MDM** | **SWLDA** | **XDAWN** | **MDM** | **SWLDA** | **XDAWN** | **SS** | **MDM** | **SWLDA** | **XDAWN** | **MDM** | **SWLDA** | **XDAWN** |
| 1 | 1 | 0.96 | 0.93 | 0.94 | 0.94 | 0.94 | 0.94 | 5 | 0.98 | 0.96 | 0.99 | 0.85 | 0.88 | 0.89 |
| 2 | 1 | 0.91 | 0.88 | 0.91 | 0.89 | 0.90 | 0.94 | 5 | 0.97 | 0.96 | 0.95 | 0.90 | 0.92 | 0.89 |
| 3 | 1 | 0.79 | 0.74 | 0.75 | 0.86 | 0.90 | 0.88 | 5 | 0.87 | 0.85 | 0.84 | 0.85 | 0.87 | 0.85 |
| 4 | 1 | 0.90 | 0.94 | 0.89 | 0.89 | 0.75 | 0.69 | 5 | 0.94 | 0.94 | 0.97 | 0.88 | 0.91 | 0.88 |
| 5 | 1 | 0.94 | 0.85 | 0.91 | 0.91 | 0.95 | 0.94 | 5 | 0.73 | 0.83 | 0.84 | 0.83 | 0.93 | 0.88 |
| 6 | 1 | 0.96 | 0.93 | 0.95 | 0.95 | 0.98 | 0.98 | 5 | 0.91 | 0.83 | 0.77 | 0.88 | 0.89 | 0.87 |
| 7 | 1 | 0.90 | 0.87 | 0.89 | 0.87 | 0.95 | 0.94 | 5 | 0.87 | 0.84 | 0.87 | 0.88 | 0.91 | 0.88 |
| 8 | 1 | 0.90 | 0.96 | 0.90 | 0.97 | 0.98 | 0.97 | 5 | 0.92 | 0.89 | 0.87 | 0.81 | 0.83 | 0.84 |
| 1 | 2 | 0.87 | 0.89 | 0.94 | 0.77 | 0.82 | 0.85 | 6 | 0.99 | 0.93 | 0.96 | 0.89 | 0.86 | 0.92 |
| 2 | 2 | 0.85 | 0.74 | 0.80 | 0.79 | 0.85 | 0.83 | 6 | 0.87 | 0.83 | 0.85 | 0.91 | 0.93 | 0.91 |
| 3 | 2 | 0.87 | 0.84 | 0.87 | 0.75 | 0.69 | 0.69 | 6 | 0.80 | 0.91 | 0.95 | 0.97 | 0.94 | 0.95 |
| 4 | 2 | 0.94 | 0.91 | 0.93 | 0.85 | 0.84 | 0.85 | 6 | 0.96 | 0.91 | 0.96 | 0.92 | 0.94 | 0.97 |
| 5 | 2 | 0.99 | 0.98 | 0.99 | 0.74 | 0.80 | 0.78 | 6 | 0.95 | 0.91 | 0.92 | 0.82 | 0.87 | 0.91 |
| 6 | 2 | 0.86 | 0.79 | 0.84 | 0.77 | 0.76 | 0.74 | 6 | 0.85 | 0.76 | 0.83 | 0.91 | 0.92 | 0.92 |
| 7 | 2 | 0.86 | 0.90 | 0.88 | 0.77 | 0.82 | 0.80 | 6 | 0.94 | 0.96 | 0.97 | 0.77 | 0.70 | 0.76 |
| 8 | 2 | 0.93 | 0.95 | 0.98 | 0.68 | 0.68 | 0.75 | 6 | 0.99 | 0.97 | 0.96 | 0.89 | 0.85 | 0.89 |
| 1 | 3 | 0.85 | 0.78 | 0.89 | 0.68 | 0.60 | 0.59 | 7 | 0.84 | 0.74 | 0.77 | 0.85 | 0.86 | 0.86 |
| 2 | 3 | 0.90 | 0.85 | 0.88 | 0.80 | 0.73 | 0.72 | 7 | 0.96 | 0.97 | 0.97 | 0.87 | 0.78 | 0.83 |
| 3 | 3 | 0.92 | 0.91 | 0.94 | 0.80 | 0.70 | 0.66 | 7 | 0.88 | 0.90 | 0.94 | 0.83 | 0.80 | 0.76 |
| 4 | 3 | 0.97 | 0.96 | 0.93 | 0.87 | 0.82 | 0.79 | 7 | 0.83 | 0.78 | 0.75 | 0.77 | 0.80 | 0.80 |
| 5 | 3 | 0.82 | 0.73 | 0.86 | 0.79 | 0.65 | 0.68 | 7 | 0.86 | 0.88 | 0.75 | 0.88 | 0.84 | 0.80 |
| 6 | 3 | 0.89 | 0.92 | 0.96 | 0.82 | 0.69 | 0.71 | 7 | 0.91 | 0.91 | 0.87 | 0.82 | 0.76 | 0.82 |
| 7 | 3 | 0.99 | 0.97 | 0.99 | 0.88 | 0.77 | 0.77 | 7 | 0.98 | 0.97 | 0.98 | 0.79 | 0.78 | 0.76 |
| 8 | 3 | 0.94 | 0.83 | 0.86 | 0.77 | 0.73 | 0.75 | 7 | 0.73 | 0.75 | 0.78 | 0.82 | 0.80 | 0.84 |
| 1 | 4 | 0.87 | 0.83 | 0.87 | 0.94 | 0.88 | 0.83 | ***M*** | ***0.90*** | ***0.88*** | ***0.89*** | ***0.84*** | ***0.83*** | ***0.83*** |
| 2 | 4 | 0.95 | 0.96 | 0.94 | 0.86 | 0.85 | 0.85 | *SD* | *0.06* | *0.07* | *0.07* | *0.06* | *0.09* | *0.09* |
| 3 | 4 | 0.94 | 0.94 | 0.90 | 0.87 | 0.86 | 0.82 | | | | | | | |
| 4 | 4 | 0.82 | 0.79 | 0.84 | 0.86 | 0.81 | 0.78 | | | | | | | |
| 5 | 4 | 0.94 | 0.91 | 0.99 | 0.83 | 0.84 | 0.81 | | | | | | | |
| 6 | 4 | 0.90 | 0.88 | 0.92 | 0.86 | 0.84 | 0.82 | | | | | | | |
| 7 | 4 | 0.83 | 0.78 | 0.81 | 0.79 | 0.79 | 0.82 | | | | | | | |
| 8 | 4 | 0.85 | 0.85 | 0.85 | 0.80 | 0.80 | 0.81 | | | | | | | |

*Online results: adaptation.*

Finally, we show the actual online results for the adaptive and non-adaptive mode of functioning. Let us remind that the adaptive and non-adaptive runs were performed in a double-blinded fashion and randomized order. At the beginning of each of its 12 levels the game Brain Invaders shows to the subject a target alien, chosen randomly among 36 aliens. After each repetition of random flashing of each alien, in such a way that each alien is flashed two times (Congedo et al., 2011), the classification algorithm destroys the alien with the highest probability of being the target based on the MDM output. If the destroyed alien is the target the subject wins the level and goes to the next level, otherwise another repetition of flashes is done. Starting from the second repetition the MDM used the cumulated distance of all repetitions to select the alien with the highest probability. Hence, the number of repetitions needed to destroy the target (NRD) is a direct measure of performance: the lower the NRD the higher the performance. The generic classifier is calibrated using online data of the preceding sessions. The individual classifier is trained in a supervised way (the labels are known) during the experiment after each repetition. Of course, the current repetition (used to select the target) is added to the training set only after the classification output is used in order to avoid biasing the results. The weights of the two classifiers (see method section) are set according to the current number of repetitions, that is, the individual classifier is weighted as alpha = min(1, Nrep/40) and the generic classifier as (1-alpha); in this way, the generic classifier is not used anymore after 40 repetitions. This value as been set arbitrarily based on pilot studies.

Figure 6 shows the mean and standard deviation NRD as a function of levels for the first session performed by all 24 subjects. As we can see, the non-adaptive MDM features a non-significant negative slope (p=0.142), meaning constant performance across levels, whereas the adaptive MDM features a significantly negative slope (p=0.02), meaning that the performance increases as the algorithm learns from the data of the subject. On the other hand, the slope of difference of the means between adaptive and non-adaptive mode is not significant. This result shows that the adaptation is effective in leading the user toward good performances.

Figure 7 shows the histogram and percent cumulative distribution of the NRD for all 24 subjects and all 12 levels of the Brain Invaders game. The cumulative distribution at the third repetition is 94.44% for the non-adaptive mode and 95.49% for the adaptive mode, that is to say, on the average of all levels and subjects about 95% of the times three or less repetitions suffice to destroy the target. These results demonstrate that our adaptive system without calibration yields performances equivalent to the traditional system with calibration, already at the first session.

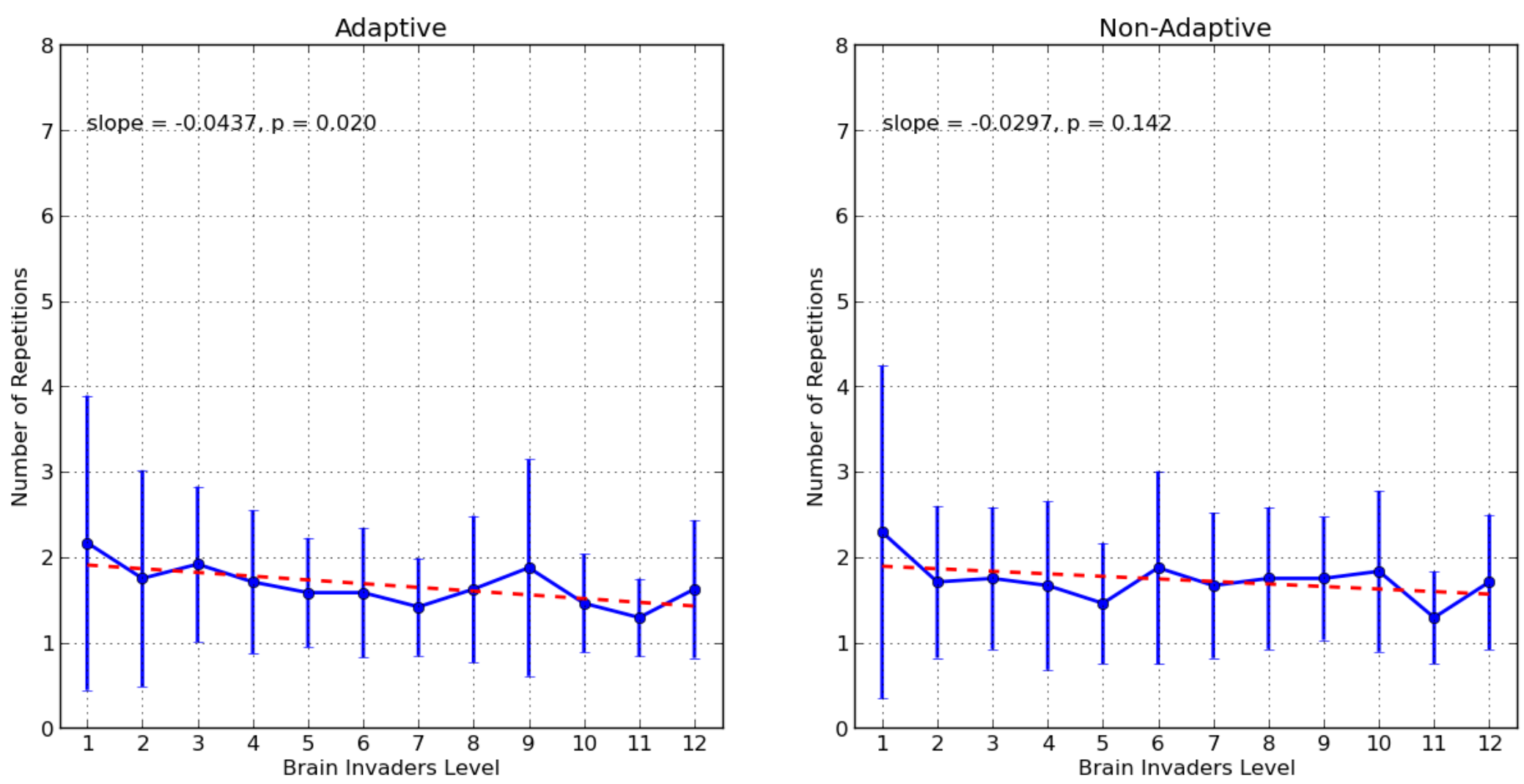

***Figure 6: Adaptation results.*** *Mean (disks) and standard deviation (bars) number of repetitions necessary for destroying the target (NRD) for the 24 subjects across the 12 levels of the first session of Brain Invaders, for the adaptive run (left) and the non-adaptive run (right). On top of the plots is printed the slope of the means and its p-value for the two-tailed test of the slope being significantly different from zero.*

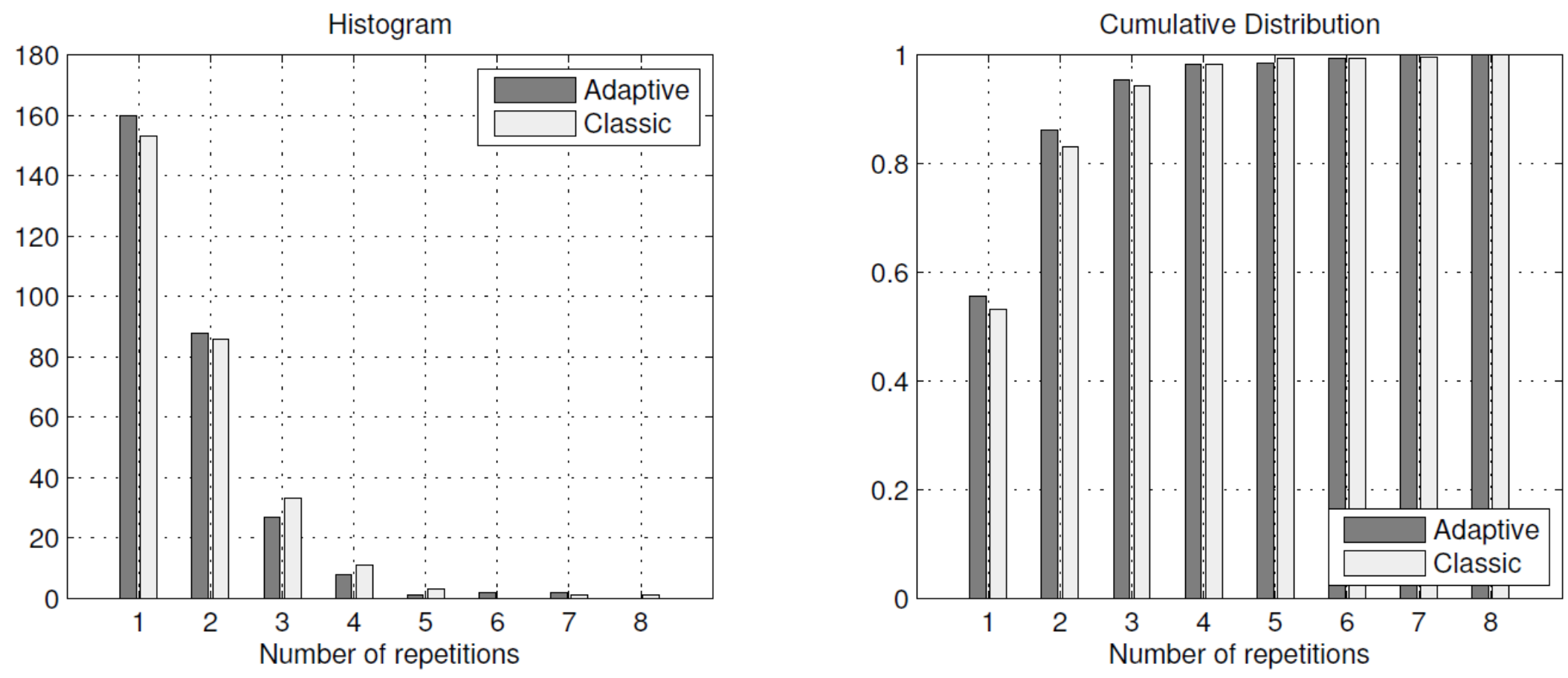

***Figure 7: Comparison of the performance of the adaptive and non-adaptive BCI as a function of the number of repetitions.*** *Raw histogram (left) and percent cumulative distribution (right) of the number of repetitions necessary to destroy the target (NRD) for all 24 subjects and all 12 levels of the first session of the Brain Invaders game.*

Figure 8 shows the means and standard deviations of the NRD for the 7 subjects across the 8 sessions of the Brain Invaders, for the adaptive runs and the non-adaptive runs. Neither slope is significantly different from zero, however the slope of the difference of the means between adaptive and non-adaptive mode is significantly smaller than zero (*slope*=-0.0304; *p*=0.047, one-tailed), demonstrating that over session the performances in the adaptive mode becomes better as compared to the non-adaptive mode. We can also appreciate the smaller standard deviation of the NRD in the adaptive mode, in all sessions. This result is striking since in non-adaptive mode the system is calibrated with data recorded just before the test.

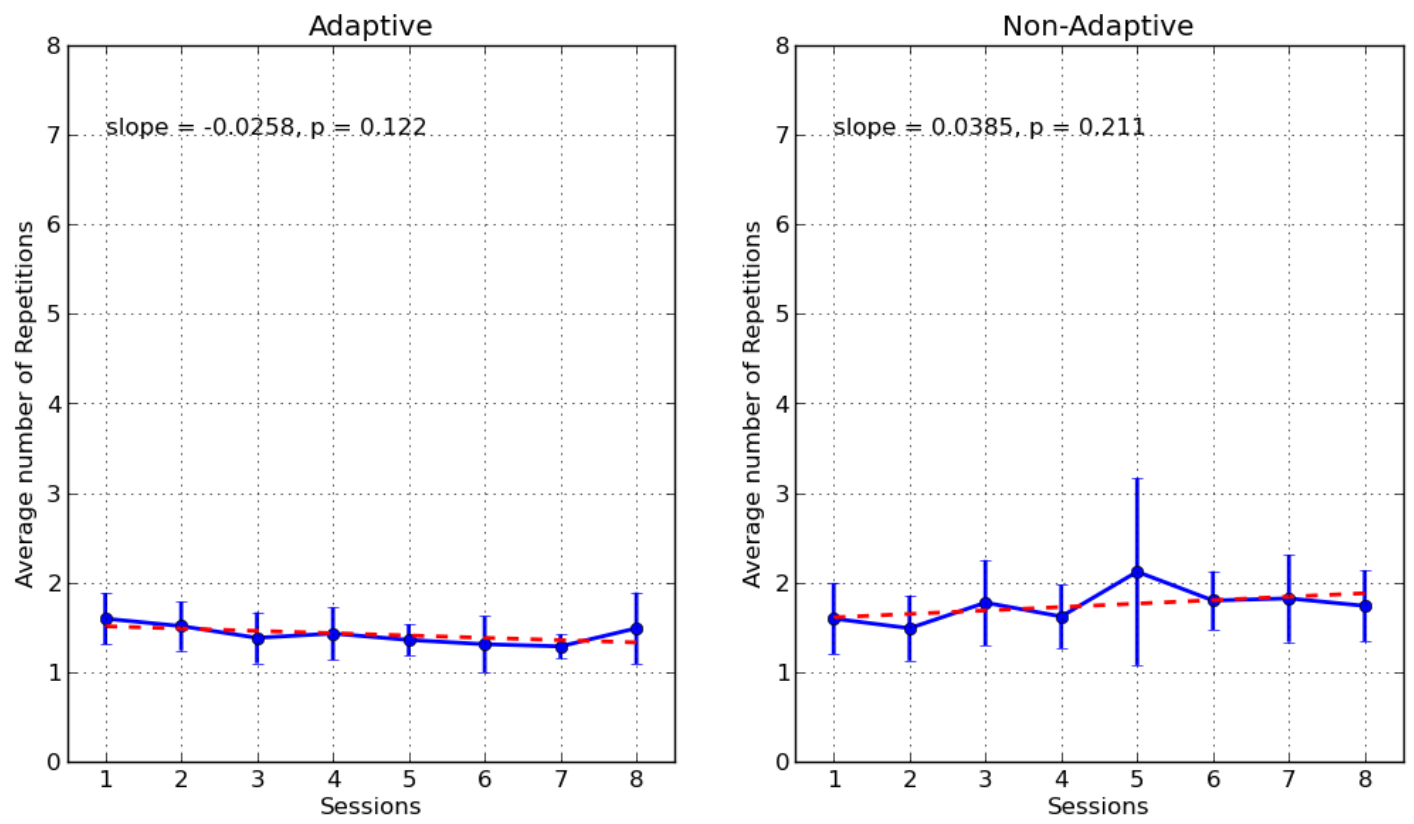

***Figure 8: Comparison of the performance of the adaptive and non-adaptive BCI over sessions.*** *Means (disks) and standard deviations (bars) of the number of repetitions necessary for destroying the target (NRD) for the 7 subjects across the 8 sessions of the Brain Invaders, for the adaptive runs (left) and the non-adaptive runs (right). On top of the plots is printed the slope of the means and its p-value for the two-tailed test of the slope being significantly different from zero.*

### *Steady-State Visually Evoked Potential Data*

Just as an example on the use of MDM for classifying SSEP data, we have applied the MDM algorithm using form of covariance matrix given by (13) to a steady-state visually evoked potential (SSVEP) dataset distributed with the OpenViBE software (Renard et *al*., 2010). The dataset is from one subject performing 32 SSVEP trials lasting six seconds. There were four classes ($Z$=4); no SSVEP (rest), 12, 15 and 20 Hz. Data was acquired by a g.tec amplifier at 512 Hz with six electrodes (CPz, O1, Oz, O2, POz, Iz). Preprocessing consisted in a 5$^{th}$ order Batterworth 2-Hz large band-pass filter centered at the three flickering frequencies.

Let us visualize the structure of the covariance matrix (13) for one subject. This illustrates well the rationale behind the choice of this form of the covariance matrix for SSEP-based BCI. See fig. 9 and its caption for details.

We applied the MDM using a 8-fold cross-validation procedure and using as data segment duration 1s, 2s, 3s, 4s, 5s and 6s. Accuracy results in term of average percent correctly classified trials are, from 1s to 6s: 53.125%, 75%, 87.5%, 93.75%, 100%, 100%, respectively.

## Conclusions

Based on the presented results we conclude that the MDM classification algorithm indeed generally do possess fast learning capabilities and that it is apt to exploit transfer learning. Also, the adaptive classification scheme we have tested has proved effective already after a few levels of the first session of the Brain Invaders; then, less than two repetitions on the average suffice to destroy the target (fig. 7), which is a remarkable result as compared to the state of the art. It should be kept in mind that this result in a fully adaptive mode has been possible thanks to the supervised mode of operation of the Brain Invaders. Even if the direct transposition of this adaptive chain to the p300 speller is possible, some difficulties remain to be addressed.

It does not matter how the covariance matrices are defined, the classification algorithms we have proposed using the Riemannian framework remains the same for all the three BCI modalities we have considered. Furthermore, it remains astonishing simple. Note that at no point there is a parameter to be tuned; it is all deterministic and completely parameter-free. This is in contrast with sophisticated machine learning approach such as support vector machine, where one or more parameter must be learned, for example, by cross-validation (Lotte et al., 2007). For this reason we claim that the strategy we have delineated is truly universal. In fact, taken together the simplicity of the MDM classification, its ability to learn rapidly (with little training data) and its good across-subject and across-session generalization, make of this strategy a very good candidate for building a new generation of BCIs.

Such BCIs will be smartly initialized using remote massive databases and will adapt to the user fast and effectively in the first minute of use. They will be reliable, robust and will maintain good performances. Having analyzed and tried several among the strategies that can be found in the literature, we believe that the Riemannian framework is a good candidate, in that it is the only one possessing all necessary properties a)-h) we have listed in the introduction.

In Barachant et *al*. (2012a) we have shown that motor imagery classification can be improved significantly over the results shown in table 1 using a Riemannian framework mapping the covariance matrices in the tangent space and applying a feature selection + LDA in the tangent space. In Barachant et *al*. (2012b) a support vector machine embedded with a Riemannian kernel was used. These two methods outperform the state of the art but they require tuning parameters. In Barachant et *al*. (2010b) we have mapped the covariance matrices in the tangent space, applied a supervised projection of the points (regularized LDA with automatic regularization) in order to increase the class separability and then remapped the data in the Riemannian manifold where the MDM applies. This method does not require tuning parameters, but it still is more involving as compared to the simple MDM. Thus, using more sophisticated classification methods in the Riemannian framework one may find a way to outperform the state of the art, but only at the expenses of the ergonomic requirements of the BCI. Similar results apply to other BCI modalities. In our view the simple MDM method is a good trade-off between accuracy, robustness and ergonomy, therefore could be considered as a starting point for the sake of a new standard suiting a large spectrum of BCI applications. Further research will find the optimal trade-off between sophistication of classification methods based on Riemannian geometry and the effectiveness/usability of the method in actual online operation. The method chosen to become a benchmark for BCI data should work without tuning parameters as MDM does and should keep the fast learning and good transfer learning capabilities.

Notice that the temporal prototypes we have defined in (5) and (9) may be defined in any plausible way, that is, they may be given by models, expectations as we do here, guesses, etc. This way to construct covariance matrices embedding both spatial and temporal information is very flexible, thus we believe it will be useful in other applications and in other domains of research.

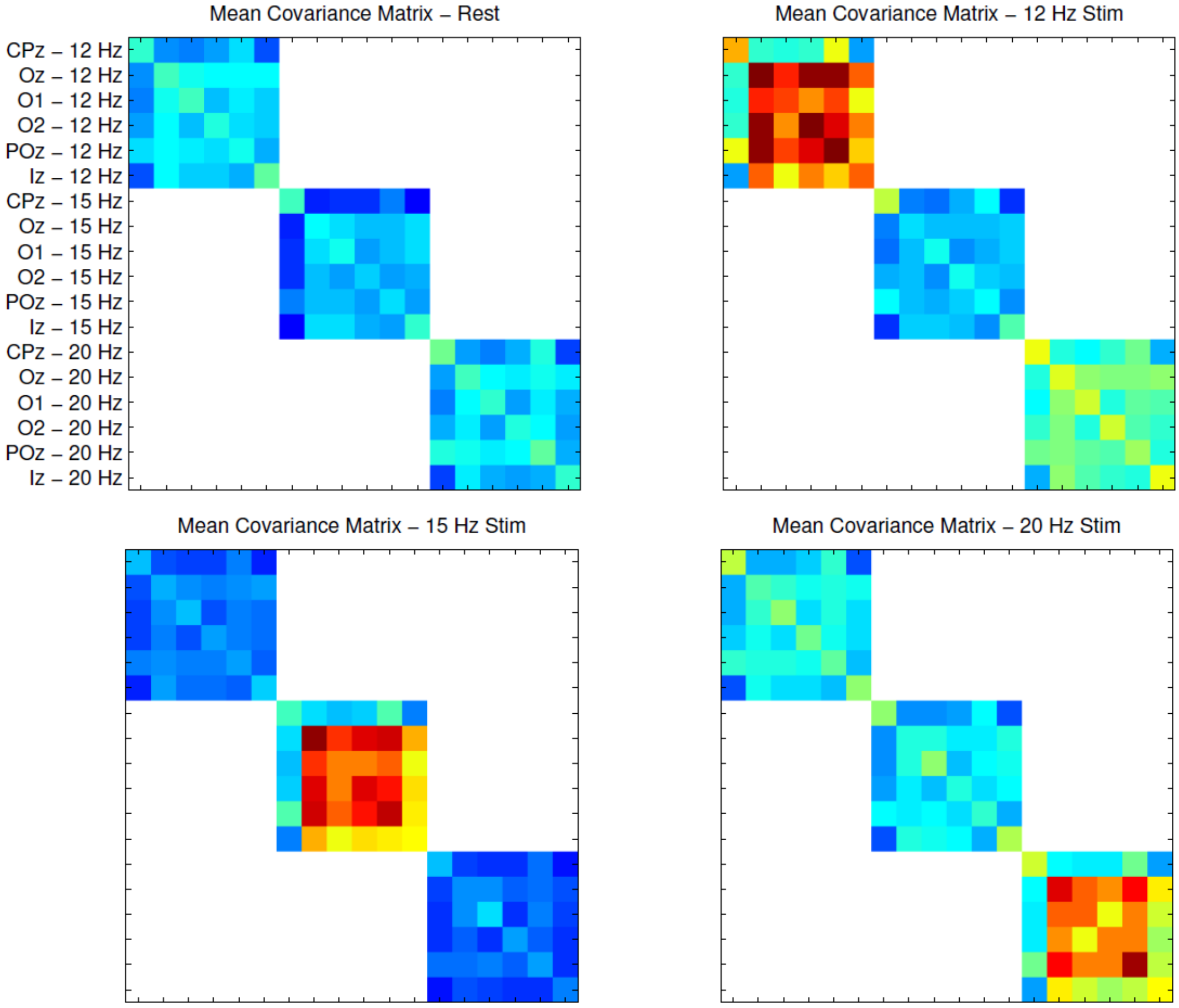

***Figure 9: Image of the covariance matrix with form* (13).** *The covariance matrix is computed on the subject performing the SSVEP experiment. Data were acquired at six electrodes. Each diagonal block of the matrices is the 6x6 covariance matrix of the data sharply band-pass filtered around the three flickering frequencies used for stimulation: 12, 15 and 20 Hz. The four matrices represented in the figure are the grand-average obtained for the "rest" class (no flickering) and for the trials with the three flickering frequencies. Notice that for each class only the block corresponding to the actual flickering frequency has high values. For the no-SSVEP (rest) data none of the blocks features high vales; this is sufficient to classify well trials belonging to any of the four classes.*

**Appendices A**

Covariance matrices are symmetric and positive-definite (SPD), implying that they can be diagonalized by a rotation (conjugation or congruence transformation by an orthogonal matrix) and have all positive eigenvalues. In symbols, let

$$\boldsymbol{C} = \boldsymbol{U}\boldsymbol{\Lambda}\boldsymbol{U}^T \tag{16}$$

be the eigenvalue-eigenvector decomposition (EVD) of SPD matrix $\boldsymbol{C}$, where $\boldsymbol{U}\in\Re^{NxN}$ is the orthogonal matrix holding in columns the eigenvectors and $\boldsymbol{\Lambda}\in\Re^{NxN}$ the diagonal matrix holding the eigenvalues (typically sorted on the diagonal in descending order of magnitude such as $\lambda_1 \geq \ldots \geq \lambda_N$ ). We introduce here the use of an exponential map, which is always SPD and induces on the space of SPD matrices an *affine-invariant metric* on a *Riemann manifold*. Such native space of SPD matrices is a regular manifold of constant curvature without boundaries, developing instead infinitely in all of its ½$N(N+1)$ dimensions (Bathia, 2013; Moakher, 2005; Pennec et *al*. 2004). In the Riemannian manifold each SPD matrix is represented by a point. In such a space the *geodesic* between two points $\boldsymbol{C}_1$ and $\boldsymbol{C}_2$ is the curve joining the two points with minimum length. Such curve is unique for a given metric. The length of the geodesic between these two points is their *distance*. Since these points are SPD matrices, the half-point on this curve, according to the chosen metric, is the *geometric mean* of the two matrices. Here and hereafter the *mean* should be understood as a *geometric* concept, not as an *arithmetic* concept. The concept of geometric mean can be extended to three of more matrices. The distance between two SPD matrices is given by (Bhatia, 2007, 2013; Moakher, 2005; Moakher and Batchelor, 2006; Pennec et *al*., 2004; Nakamura, 2009; Sra, 2012)

$$\delta_R\left(\boldsymbol{C}_1 \leftrightarrow \boldsymbol{C}_2\right) = \left\|\ln\left(\boldsymbol{C}_1^{-\frac{1}{2}}\boldsymbol{C}_2\boldsymbol{C}_1^{-\frac{1}{2}}\right)\right\|_F = \sqrt{\sum\nolimits_n \left[\ln\left(w_n\right)\right]^2}\ , \tag{17}$$

where $\left\|\cdot\right\|_F$ is the Frobenius norm and $w_1,\ldots,w_N$ are the eigenvalues of either $\boldsymbol{C}_1^{-1}\boldsymbol{C}_2$ or $\boldsymbol{C}_2^{-1}\boldsymbol{C}_1$.

The geometric mean between $K$>2 SPD matrices $\boldsymbol{C}_k\in\{\boldsymbol{C}_1,\ldots,\boldsymbol{C}_K\}$ can be found by a globally convergent iterative algorithm (Pennec et *al*., 2004; Manton, 2004):

| Geometric Mean $\boldsymbol{M}$ of $K$ SPD matrices $\boldsymbol{C}_k\in\{\boldsymbol{C}_1,\ldots,\boldsymbol{C}_K\}$ (18) |
| --- |
| *Initialize* $\boldsymbol{M}$ *by a smart guess or by the arithmetic mean* $\frac{1}{K}\sum_k \boldsymbol{C}_k$ . |
| *Repeat* $\boldsymbol{M} \leftarrow \boldsymbol{M}^{\frac{1}{2}} exp\left[\frac{1}{K}\sum_k ln\left(\boldsymbol{M}^{-\frac{1}{2}}\boldsymbol{C}_k\boldsymbol{M}^{-\frac{1}{2}}\right)\right]\boldsymbol{M}^{\frac{1}{2}}$ |
| *until the Frobenius norm of* $\sum_k ln\left(\boldsymbol{M}^{-\frac{1}{2}}\boldsymbol{C}_k\boldsymbol{M}^{-\frac{1}{2}}\right)$ *is small enough.* |

For (17) and (18) here above we have been using the following functions of eigenvalues, which are readily derived from the EVD (16):

$$\boldsymbol{C}^{-1} = \boldsymbol{U\Lambda}^{-1}\boldsymbol{U}^{T} \ ; \ \boldsymbol{C}^{\frac{1}{2}} = \boldsymbol{U\Lambda}^{\frac{1}{2}}\boldsymbol{U}^{T} \ ; \ \boldsymbol{C}^{-\frac{1}{2}} = \boldsymbol{U\Lambda}^{-\frac{1}{2}}\boldsymbol{U}^{T} \ ; exp\left(\boldsymbol{C}\right) = \boldsymbol{U}exp(\boldsymbol{\Lambda})\boldsymbol{U}^{T} \ ; \ ln\left(\boldsymbol{C}\right) = \boldsymbol{U}ln(\boldsymbol{\Lambda})\boldsymbol{U}^{T} \ .$$